\documentclass[aps,amssymb,floatfix,prd,twocolumn, preprintnumbers]{revtex4}
\usepackage[utf8]{inputenc}
\usepackage{epstopdf}
\usepackage{capt-of}
\usepackage{graphicx}  % Include figure files
\usepackage{dcolumn}   % Align table columns on decimal point
\usepackage{bm}% bold math
\usepackage{amsmath}
\usepackage[font=scriptsize]{caption}
\usepackage[colorlinks]{hyperref}
\usepackage{xcolor}
\usepackage{orcidlink}
\begin{document}
\input epsf.tex
%%%%%%%%%%%%
%\title{Cosmic Dynamics and Energy Conditions in a Reconstructed $f(T)$ Gravity Accelerating Model with a Parametrized Hubble Function}
%\title{Reconstructed $f(T)$ Gravity Accelerating Model with Parametrized Hubble Function}
\title{Reconstructing $f(T)$ Gravity From Hubble Parameterization Constraints}
\author{Suraj Kumar Behera \orcidlink{0009-0009-9294-460X}}
\email{skbehera.researches@gmail.com}
\affiliation{Department of Mathematics (School of Advanced Sciences), Vellore Institute of Technology - Andhra Pradesh University, Amaravati, Guntur District, Andhra Pradesh 522241, India}

\author{Pratik P. Ray \orcidlink{0000-0003-2304-0323}}
\email{pratik.chika9876@gmail.com}
\affiliation{Department of Mathematics (School of Advanced Sciences), Vellore Institute of Technology - Andhra Pradesh University, Amaravati, Guntur District, Andhra Pradesh 522241, India}
\affiliation{Pacif Institute of Cosmology and Selfology (PICS), Sagara, Sambalpur 768224, Odisha, India}

\author{B. Mishra\orcidlink{0000-0001-5527-3565}}
\email{bivu@hyderabad.bits-pilani.ac.in}
\affiliation{Department of Mathematics,
Birla Institute of Technology and Science-Pilani, Hyderabad Campus, Jawahar Nagar, Kapra Mandal, Medchal District, Telangana 500078, India.}

%%%%%%%%%%%%

\affiliation{}

\begin{abstract}
\begin{center}
{\textbf{Abstract}}
\end{center} 
In this paper, we have presented the cosmological model of the Universe that represents late time cosmic acceleration in torsion based gravitational theory, the $f(T)$ gravity. A well motivated parametrization for the Hubble parameter has been introduced and the free parameters involved are constrained using the cosmological datasets. With the constrained values of the free parameters, other geometrical parameters such as deceleration parameter, jerk parameter, and snap parameter are analyzed and confronted with the prescribed value of the cosmological observations. In addition, the dynamical parameters are analyzed in some non-linear form of $f(T)$ and the energy conditions are also studied and confirmed with the violation of the strong energy condition. The obtained cosmological model provides late time phantom behavior of the Universe. 
\end{abstract}

\maketitle
%\textbf{PACS number

\section{Introduction} \label{introduction}
Our understanding on the evolutionary behavior of the Universe has changed after the revelation of accelerated expansion behavior of the Universe\cite{Riess_1998, Perlmutter_1999}. The first impression was that an exotic form of energy known as dark energy(DE) is responsible for such a strange behavior. General Relativity(GR) has been instrumental in addressing many cosmic issues, in particular the solar system test \cite{misner1973gravitation,capozziello2011extended,faraoni2008f}, but has limitations in identifying the reason behind the present behavior of the Universe. The first attempt made in this direction was the introduction of the cosmological constant $(\Lambda)$\cite{sahni2000case}, but from the theoretical point of view, it lacks clarity. So, modification has been inevitable to address the late time cosmic acceleration issue. There are three equivalent geometrical approaches to understand the gravity such as curvature, torsion and nonmetricity. In the torsion approach, the teleparallel equivalent of GR (TEGR) has been introduced \cite{moller1961conservation,pellegrini1963tetrad,hayashi1979new}. In the context  of Riemannian geometry, the torsion tensor is severely limited. In particular, because of the symmetry of the christoffel's symbols, the torsion tensor is limited to zero, that is, $T^{\mu}_{\nu \lambda} = 0$. In an intriguing extension of Riemannian geometry, in the witzenb\"{o}ck space \cite{weitzenböck1923invarianten}, the Riemannian curvature tensor is zero, and the torsion tensor is non-zero ($T^{\mu}_{\nu \lambda} = 0$). This leads to a flat spacetime with the significant characteristic of absolute parallelism, also referred to as teleparallelism. Einstein pioneered the use of witzenb\"ock-type spacetime in physics by developing a unified teleparallel theory that harmonized gravity and electromagnetism \cite{einstein2005riemann}. Teleparallel gravity(TG) utilizes tetrad fields $e^i_{\mu}$ instead of metric tensor $g_{\mu \nu}$ as the fundamental variables defining the gravitational characteristics of spacetime. Instead of curvature, torsion, the TG originates from the tetrad fields and can be used as a general description of the gravitational effect. 

The first modification to TEGR is the $f(T)$ gravity theory \cite{PhysRevD.79.124019,Linder_2010}, where the torsion scalar $T$ has been replaced with $T+f(T)$ in the action. The advantage of $f(T)$ gravity is that its field equations are of second-order as compared to $f(R)$ gravity, which has fourth-order field equations \cite{aldrovandi2012teleparallel}. The applications of $f(T)$ gravity theory have been extensively investigated in the study of astrophysical and cosmological phenomena. Significantly, such theories are frequently used to propose an alternate explanation for large-scale structure, specifically the late-time accelerating expansion of the Universe, without invoking the DE \cite{PhysRevD.83.104017,Myrzakulov_2011,Cai_2016,PhysRevD.75.084031,PhysRevD.78.124019,bengochea2009dark,PhysRevD.81.127301,PhysRevD.85.044033,PhysRevD.89.124036,Harko_2014,BAHAMONDE201737,Capozziello_2017,PhysRevD.97.124064,Awad_2018,Jiménez_2018,Golovnev_2018,PhysRevD.98.124013,PhysRevD.99.064006,universe5030080,universe5060139,universe5060143}. We will discuss some of the recent studies in the framework of $f(T)$ gravity. In spite of proposing conformal scale and gauge field theories and building conformal torsion gravity, Bamba et al. \cite{PhysRevD.88.084042} have investigated several conformal difficulties of pure and extended TG. Using several kinds of data sources, including the Pantheon supernovae sample, Hubble constant measurements, cosmic microwave shift parameter, and redshift-space distortion measurement, the $f(T)$ gravity has been constrained both at background and perturbation levels in Ref. \cite{PhysRevD.100.083517}. The quasinormal mode frequencies of a test massless scalar field around static black hole solutions in $f(T)$ gravity have been determined by by Zhao et al. \cite{Zhao_2022}. Using various kinds of $f(T)$ gravity models, Jackson et al. \cite{Said_2020} have investigated the direct correlation between a temporal violation of the electromagnetic fine-structure constant and the violation of the distance-duality relation. Rezaei and Amani \cite{REZAEI20171068} examined the stability of extended $f(T)$ gravity with an energy-momentum tensor coupling within the framework of a modified Chaplygin gas.

Recent research have expanded the teleparallel framework via generalized $f(T)$ gravity theories, investigating their cosmological consequences and aligning with observations. Mishra \textit{et al.}\cite{mishra2025gaugeinvariantperturbationsfttg} presented a gauge-invariant cosmological perturbation framework for $F(T,T_G)$ gravity, exploring scalar, vector, and tensor modes to determine stability conditions and validate consistency with observational constraints. Late-time quintessence cosmologies governed by power-law, hyperbolic, and axion-like scalar field potentials with constraining of model parameters were explored in Ref. \cite{duchaniya2025quintessencemodelslateuniverse}. Lohakare \textit{et al.}\cite{LOHAKARE2023101164} examined the $F(T,T_G)$ gravity model using Hubble and Pantheon data, indicating a viable transition decelerated phase to accelerated expansion in accordance with observations. The Noether symmetry approach in $f(T,T_G)$ cosmology \cite{Kadam_2023}, yielding exact solutions that describe de sitter and accelerating phases of the Universe. A model-independent approach for inflation in scalar-tensor gravity was established by incorporating a power-law non minimal coupling between scalar field and torsion, exhibiting its consistency with observational constraints and sub-luminal scalar perturbations\cite{FOMIN2025101895}. An investigation of dynamical systems of different $f(T)$ gravity models was conducted at both background and perturbation levels, demonstrating stable critical points and late-time acceleration consistent with current cosmological observations\cite{DUCHANIYA2024101461}. The propagation of gravitational waves in teleparallel Gauss–Bonnet gravity was explored by deriving the tensor perturbation equations, illustrating that gravitational waves propagate at the speed of light and comply with existing observational constraints from multi-messenger astronomy \cite{p6l8-k963}.

Based on the success of TG in addressing the present cosmic issue, we are motivated to explore a well motivated functional form of $f(T)$ to address late time cosmic phenomena in $f(T)$ gravity. We employ the simple parameterization of a particular form of the Hubble parameter using Hubble, BAO, Pantheon+SH0ES and Hubble + BAO + Pantheon+SH0ES observational datasets and generate solutions for the modified Friedmann equations in the FLRW spacetime. The structure of the paper is as follows: in section \ref{field equations}, we have discussed TG and the field equations of $f(T)$ gravity. In section \ref{reconstruction}, a specific form of the Hubble parameter is discussed and parametrized using $H(z)$, BAO, Pantheon+SH0ES and combination of all three datasets via MCMC approach. In section \ref{model}, the dynamical parameters are analyzed with some specific form of $f(T)$ and the energy conditions are presented. In sections \ref{conclusion}, the summary and conclusions of the cosmological models are discussed.
     
\section{Field Equations of $f(T)$ Gravity} \label{field equations}
In TG, tetrads are employed as the dynamical variable and it must satisfy the orthogonality condition,
\begin{equation}
e^i_{\mu} e^{\mu}_j = \delta^i_j  ~~~~~~~~~~~~~~~~~~~~~~~~~~~~~~~~~~~~e^{\mu}_{i} e^{i}_{\mu} = \delta^{\mu}_{\nu},
\end{equation}
where $e^{\mu}_{i}$ represents the inverse tetrad. The tetrad and metric tensor $g_{ij}$ are related through Minkowski space time as,
\begin{equation}
g_{\mu \nu}=\eta_{ij}e^i_{\mu} e^j_{\nu},
\end{equation}

where, $\eta_{ij} = diag (1,-1,-1,-1)$ represents the Minkowski metric. One can relate the Witzenb{\"o}ck and TG connection \cite{weitzenbock1923invariantentheorie} with the following expression, 

\begin{equation}
\Gamma^{\lambda}_{\nu \mu} \equiv e^{\lambda}_i \partial_{\mu} e^i_{\nu}.
\end{equation}
The torsion tensor can be expressed as,
\begin{equation}
T^{\lambda}_{\mu \nu} = \hat{\Gamma}^\lambda_{\ \nu\mu} - \hat{\Gamma}^\lambda_{\ \mu\nu}
= e^{\lambda}_i (\partial_{\mu} e^i_{\nu} - \partial_{\nu} e^i_{\mu}). 
\end{equation}
Further the contortion tensor and superpotential tensor can be expressed respectively as
\begin{eqnarray}
    S^{\;\;\;\mu \nu}_{\rho} &\equiv& \frac{1}{2} (K^{\mu \nu}_{\;\;\;\rho} + \delta^{\mu}_{\rho}T^{\alpha \nu}_{\;\;\;\;\alpha} -\delta^{\nu}_{\rho}T^{\alpha \mu}_{\;\;\;\;\;\alpha}), \label{superpotential}\\
    K^{\mu \nu}_{\;\;\;\;\rho} &\equiv&  \frac{1}{2} (T^{\nu \mu}_{\;\;\;\;\rho}+T^{\;\;\;\mu \nu}_{\rho}-T^{\mu \nu}_{\;\;\;\;\rho})\label{contortion}.
\end{eqnarray}
The torsion scalar $T$ can be obtained by contracting the torsion tensor as,
\begin{equation}
    T = S^{\mu \nu}_{\alpha} T^{\alpha}_{\mu \nu} = \frac{1}{2} T^{\alpha \mu \nu} T_{\alpha \mu \nu} + \frac{1}{2} T^{\alpha \mu \nu} T_{\nu \mu \alpha} - T^{\alpha}_{\alpha \mu} T^{\nu \mu}_{\nu} \label{torsionscalar}
\end{equation}
The action of $f(T)$ gravity \cite{Cai_2016} is,

\begin{equation}
S=\frac{1}{16 \pi G}\int d^4xe\left[T+f\left(T\right)+\mathcal{L}_m\right], \label{action}
\end{equation}
where $\mathcal{L}_m$ is the total matter Lagrangian, $e = det[e^i_{\mu}] = \sqrt{-g}$, and $G$ is the gravitational constant. The natural system, $\kappa^2 = 8 \pi G = 1$, has been considered. The gravitational field equations can be obtained by varying the action \eqref{action} with respect to the vierbein as,

\begin{multline}
e^{-1}\partial_{\mu}\!\left( e\, e^{\rho}_i S^{\;\;\mu \nu}_{\rho} \right)\!\left[ 1 + f_T \right]
+ e^{\rho}_i S^{\;\;\mu \nu}_{\rho}\,\partial_{\mu}(T) f_{TT} \\
- e^{\lambda}_i T^{\rho}_{\;\;\mu \lambda} S^{\;\;\nu \mu}_{\rho}\!\left[ 1 + f_T \right]
+ \frac{1}{4} e^{\nu}_i \big[ T + f(T) \big]
= 4\pi G\, e^{\rho}_a T^{\;\;\nu}_{\rho}.
\label{fieldequations}
\end{multline}

For brevity, we denote $f\equiv f(T)$, $f_T=\frac{df}{dT}$, $f_{TT}=\frac{d^2f}{dT^2}$ and the total energy momentum tensor  as $T^{\;\;\nu}_{\rho}$. Now, the field equation of $f(T)$ gravity in flat FLRW spacetime
\begin{equation}\label{spacetime}
ds^2=dt^2-a^2 (t) \delta_{ij}dx^i dx^j,
\end{equation}
where $a(t)$ be the scale factor and $i,j=0,1,2,3$ can be obtained as
\begin{eqnarray}
3H^2 &=& 8 \pi G \rho_m-\frac{f}{2}+T f_T, \label{energydensity}\\
\dot{H} &=& -\frac{4 \pi G (\rho_m + p_m)}{1 + f_T + 2 T f_{TT}}.\label{pressure}
\end{eqnarray}
In the FLRW scenario, the torsion scalar becomes $T=-6H^2$. The Hubble parameter, $H \equiv \frac{\dot{a}}{a}$, where an over dot indicates ordinary derivative with respect to cosmic time $t$. The matter sector comprises the energy density $\rho_m$ and pressure $p_m$, Now, one can express the dark energy density $(\rho_{de})$ and the dark energy pressure $(p_{de})$ as,

\begin{eqnarray}
\rho_{de}&=& \frac{1}{16 \pi G} [-f + 2 T f_T],\label{rhode}\\
p_{de} &=& -\frac{1}{16 \pi G} \left[\frac{-f + T f_T - 2 T^2 f_{TT}}{1 + f_T + 2 T f_{TT}}\right].\label{pde}
\end{eqnarray}
and subsequently the dark energy equation of state (EoS) parameter $(\omega_{de}=\frac{p_{de}}{\rho_{de}})$ can be obtained as,
\begin{equation}\label{omegade}
\omega_{de} = -1 + \frac{(f_T + 2 T f_{TT}) (-f + T + 2 T f_T)}{(1 + f_T + 2 T f_{TT}) (-f + 2 T f_T)}.
\end{equation}
We intend to study the dynamical behavior of the cosmological model, hence we incorporate some well-motivated parametric form of Hubble Parameter and constrain all the model parameters via Bayesian statistics. 

\section{ Reconstruction of $H(z)$ from observational data } \label{reconstruction}
Hubble parameter measures the rate of expansion of the Universe and is one of the key components in any cosmological model framed in gravitational theories. However, one can follow a model-independent approach to explore cosmic evolution without assuming any gravitational theory \cite{PhysRevD.87.023520}. Here, we employ a model-independent technique to solve the field equations using the cosmological parameterization, in fact, the three unknowns in the field equations such as the Hubble parameter, pressure, and energy density. The shift from decelerating to accelerating evolutionary behavior of the Universe involves the parameterization of Hubble parameter, deceleration parameter and EoS parameter and can be validated through the observational datasets. In such approach, the late time cosmic acceleration has been shown by introducing some functional form of $H(z)$ \cite{doi:10.1142/S0219887817501110,MYRZAKULOV2023345,YADAV2024114}. With this motivation we express the parametric form of Hubble parameter as 
\begin{equation} \label{HP}
H(z)=H_0 \sqrt{(z-a) (b z+1)+(a+1)},
\end{equation}
where $H_0$ denotes the present value of Hubble parameter, $a$ and $b$ are free parameters which would be constrained using the cosmological datasets. The deceleration parameter ($q$) decides the accelerating ($q<0$) or decelerating ($q>0$) behavior of the cosmological model and can be expressed in terms of Hubble parameter as, 

\begin{equation} \label{DP}
q(z) = - \frac{\dot{H}}{H^2} - 1.
\end{equation}
Using the expression $\dot{H} = - (z+1) H(z) \frac{dH}{dz}$ and Eqn. \eqref{HP}, Eqn. \eqref{DP} can be written in redshift as,

\begin{equation}\label{DP1}
q (z) = -\frac{a (b - b z) - 2 b z + z + 1}{2 (-a b z + b z^2 + z + 1)}
\end{equation}
This parameterization offers new view points on the evolution of cosmic acceleration and its observational implications.\\
Now, we shall use observational data from several cosmological surveys to constrain the free parameters of $H(z)$, i.e., by investigating data that describe the distance-redshift relation. For this, we used expansion rate measurements from early-type galaxies, such as $H(z)$ data, Baryon Acoustic Oscillations (BAO) data, Pantheon+SH0ES data and a combination of all. These observational datasets are essential for estimation of cosmological parameter and are independent of any particular cosmological model. A brief description of each dataset is given below.\\
 
\noindent{\bf{ The Hubble data}}: It has 32 data points for the Hubble parameter covering the redshift range $ 0.07 \leq z \leq 1.965 $ \cite{Zhang_2014}. The Chi-square $(\chi^2)$ function is minimized to obtain the best-fit values of the model parameters $H_0$, $a$, and $b$. The Chi-square function is defined as,
\begin{equation}
\chi^2_{H(z)} = \sum_{i=1}^{32} \left[ \frac{\left(H_\text{th}(z_i, \psi) - H_\text{obs}(z_i)\right)^2}{\sigma_H^2(z_i)} \right],
\end{equation}
where $\psi$ is the vector of the cosmological background parameters. $H_\text{th}(z_i, \psi)$ and $H_\text{obs}(z_i)$ respectively refers to developed and observed Hubble parameter. The observational errors in the observed value $H_\text{obs}(z_i)$ denoted as $\sigma_H^2(z_i)$.\\

\noindent{\bf{The BAO data}}: A vital cosmological probe for investigating the large-scale structure of the Universe is the Baryon Acoustic oscillations(BAO). At early times, baryon matter and radiation are compressed by acoustic waves within the photon-baryon fluid, resulting in these oscillations. A standard rule for measuring cosmic distances can be created by this compression, which results in a noticeable peak in the correlation function of galaxies of quasars. The sound horizon at the epoch of recombination determines the comoving scale of the BAO peak and is dependent on the cosmic microwave background (CMB) temperature and the baryon density. The location of the BAO peak in the angular direction at a specific redshift $z$ defines the angular separation as,

\begin{equation}
\Delta\theta = \frac{r_d}{(1 + z) D_A(z)},
\end{equation}
where $D_A (z)$ is the angular distance. In the radial direction, the redshift separation is defined as,

\begin{equation}
\Delta z = \frac{r_d}{D_H(z)},
\end{equation}
where the Hubble distance by $D_H = c/H$ and the sound horizon at the drag epoch by $r_d$. We can constrain cosmological parameter combinations that define $D_H/r_d$ and $D_A/r_d$ by accurately measuring the BAO peak position at various redshifts. $H(z)$ can also be estimated by choosing a suitable value for $r_d$. Here we use a dataset of 26 independent data points derived from line-of-sight BAO measurements \cite{10.1111/j.1365-2966.2009.15405.x}. The $\chi^2$ function for the BAO data set is defined as,

\begin{equation}
\chi^2_{B A O}(\phi) = \sum_{i=1}^{26} \left[ \frac{(H_{th}(z_i, \phi) - H_{obs}(z_i))^2}{\sigma^2_H(z_i)} \right],
\end{equation}
where $H_{th}(z_i, \phi)$ represents the theoretical value of the Hubble, the model parameters are represented by $\phi$, $H_{obs}(z_i)$ represents the observed value of the Hubble from the BAO analysis, and the related error in the observed BAO data points is denoted by $\sigma_H(z_i)$.\\

\begin{figure}[htbp] 
   \centering 
   \mbox{\includegraphics[scale=0.45]{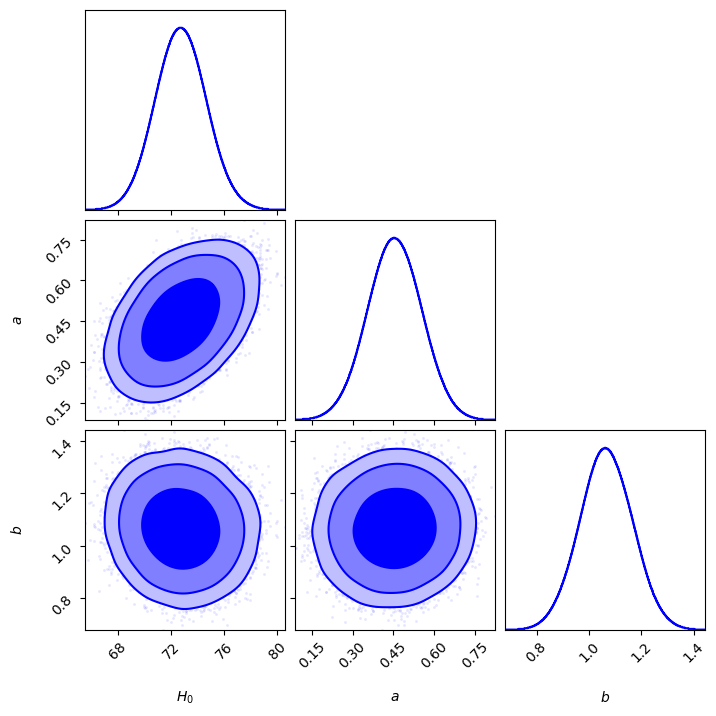}}   
    \caption{\raggedright Two dimensional contour diagram extracted from $H(z)$ data demonstrate the favored parameter ranges and uncertainty contours (up to $3 \sigma$) for $H_0$, $a$, and $b$.} 
   \label{MCMC with Hubble dataset}
\end{figure}

\begin{figure}[htbp] \label{bao}
   \centering 
   \mbox{\includegraphics[scale=0.45]{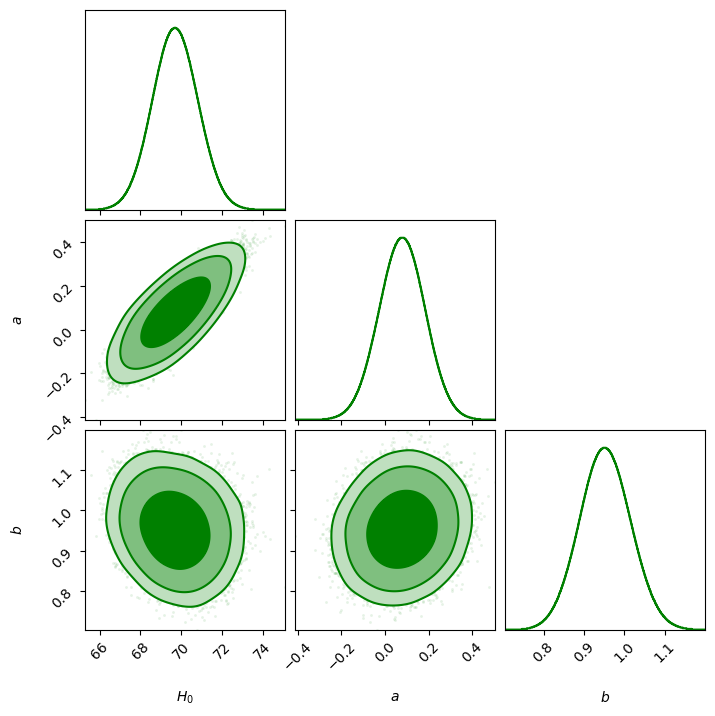}}   
    %\hspace{10px}
    %\mbox{\includegraphics[scale=0.53]{model 1_1.pdf}}
    %\hspace{10px}
    %\mbox{\includegraphics[scale=0.53]{model 1_2.pdf}}
    %\hspace{10px}
    %\mbox{\includegraphics[scale=0.53]{model 1_3.pdf}}
    \caption{\raggedright Two dimensional contour diagram extracted from BAO data demonstrate the favored parameter ranges and uncertainty contours (up to $3 \sigma$) for $H_0$, $a$, and $b$.}
   \label{MCMC with BAO dataset}
\end{figure}

\begin{figure}[htbp] \label{pantheonshoes}
   \centering 
   \mbox{\includegraphics[scale=0.45]{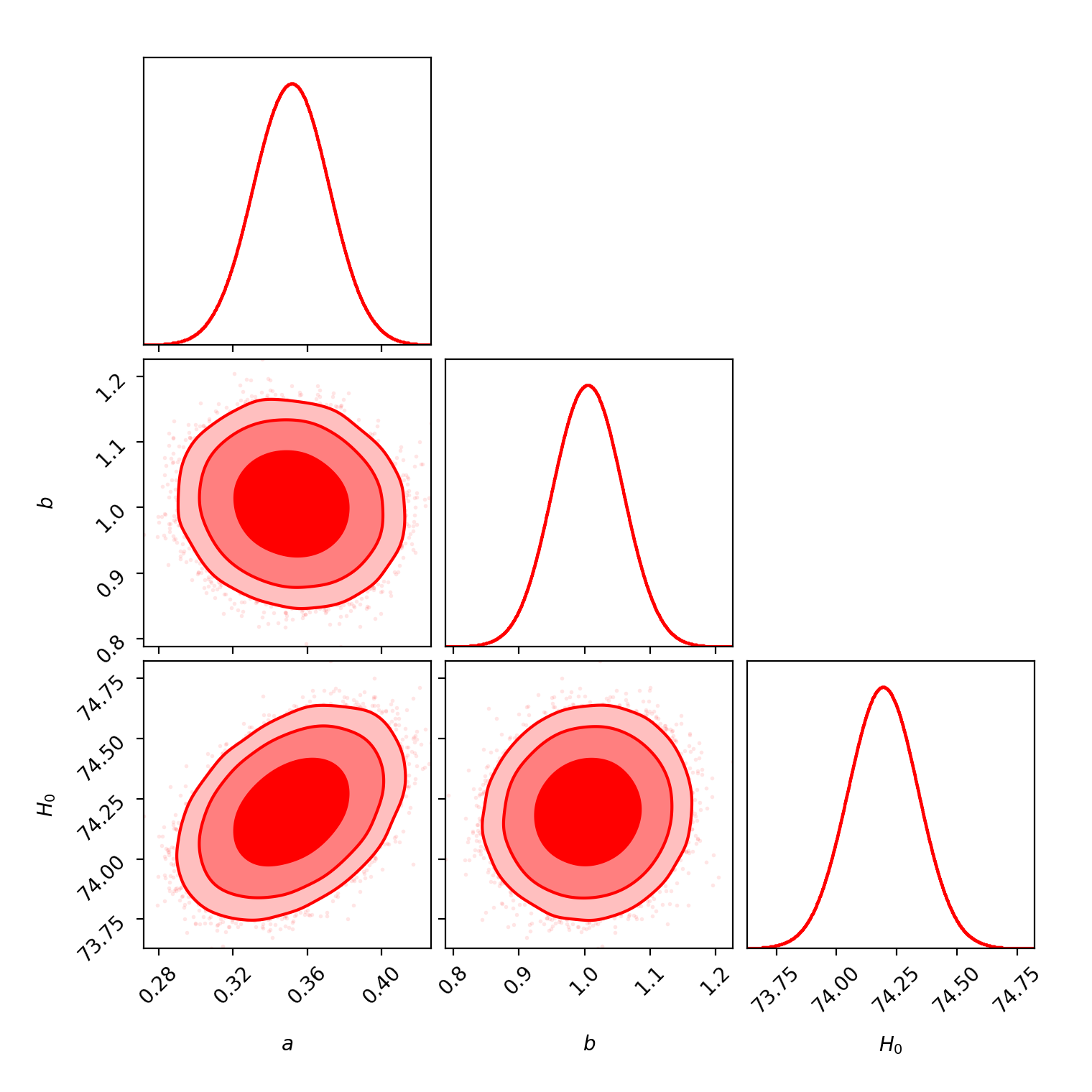}}   
    %\hspace{10px}
    %\mbox{\includegraphics[scale=0.53]{model 1_1.pdf}}
    %\hspace{10px}
    %\mbox{\includegraphics[scale=0.53]{model 1_2.pdf}}
    %\hspace{10px}
    %\mbox{\includegraphics[scale=0.53]{model 1_3.pdf}}
    \caption{\raggedright Two dimensional contour diagram extracted from Pantheon+SH0ES data demonstrate the favored parameter ranges and uncertainty contours (up to $3 \sigma$) for $H_0$, $a$, and $b$.}
   \label{MCMC with Pantheon+SH0ES dataset}
\end{figure}

\begin{figure}[htbp] \label{combined}
   \centering 
   \mbox{\includegraphics[scale=0.45]{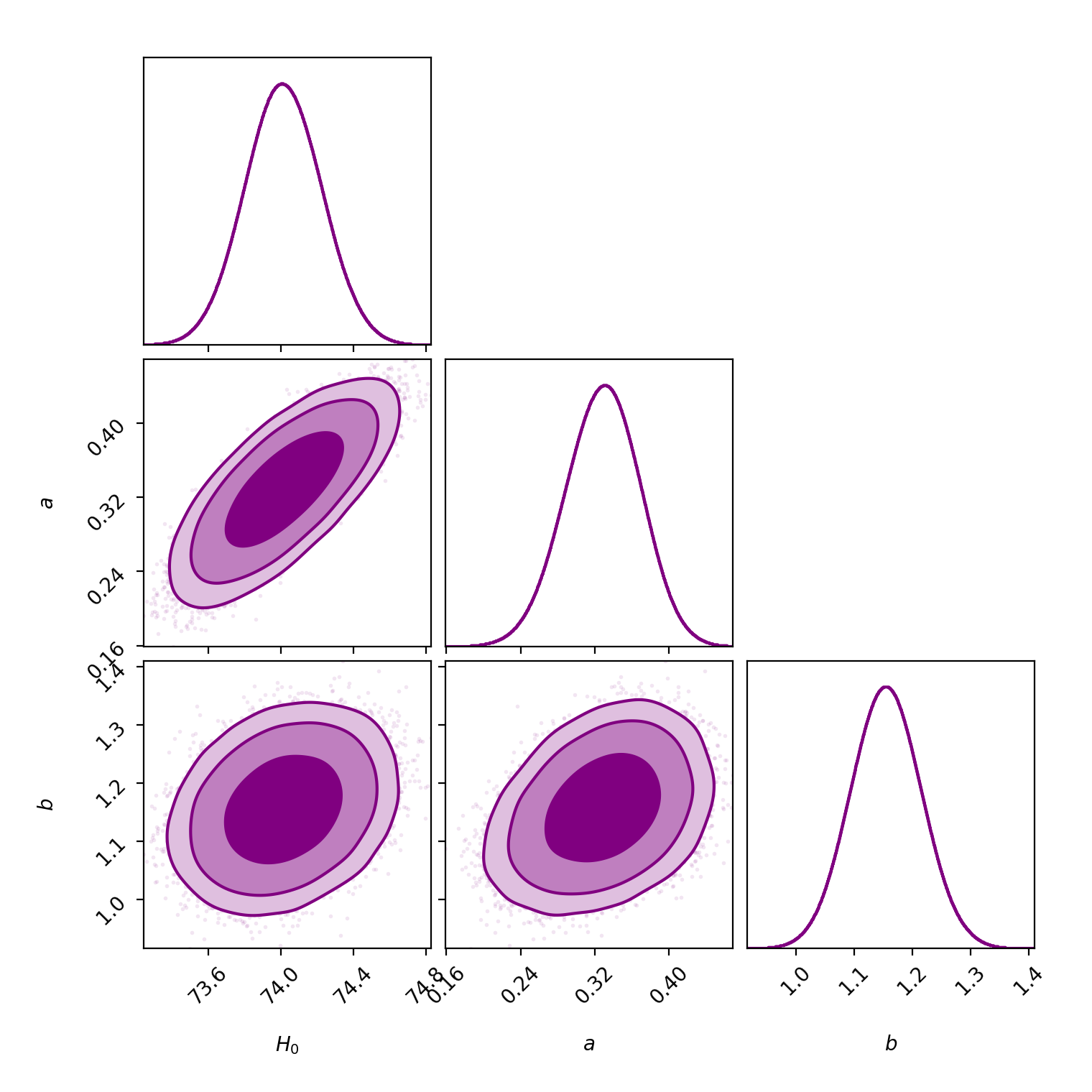}}   
    %\hspace{10px}
    %\mbox{\includegraphics[scale=0.53]{model 1_1.pdf}}
    %\hspace{10px}
    %\mbox{\includegraphics[scale=0.53]{model 1_2.pdf}}
    %\hspace{10px}
    %\mbox{\includegraphics[scale=0.53]{model 1_3.pdf}}
    \caption{\raggedright Two dimensional contour diagram extracted from combined ($H(z)$ + BAO + Pantheon+SH0ES) data demonstrate the favored parameter ranges and uncertainty contours (up to $3 \sigma$) for $H_0$, $a$, and $b$.}
   \label{MCMC with combined dataset}
\end{figure}

\noindent{\bf Pantheon+SH0ES Data}: Type Ia Supernovae (SNIa) serve as a highly reliable cosmic probe of the late universe. Their incredible uniformity in peak luminosity allows them to serve as standard candles, facilitating precise measurements of luminosity distances over a wide redshift range. This capability has played a crucial role in establishing the accelerated cosmic expansion. In recent decades, several extensive compilations of SNIa have been developed, including Union, union 2.1, JLA, Pantheon, and the most recent is Pantheon+SH0ES sample \cite{Kowalski_2008}. This sample consists of 1701 light curves corresponding to 1550 supernovae, spanning the redshift range of $0.001<z<2.26$. The theoretical distance modulus can be defined as

\begin{equation}
    \mu_{\text{th}}(z, \theta) = 5 \log_{10}\!\left( d_L(z, \theta) \right) + 25,
\end{equation}
where $\theta=\left[H_0,a,b\right]$ represents the set of Hubble model parameters. The luminosity distance can be expressed as,

\begin{equation}
d_{L}(z, \theta) = (1+z)\,c \int_{0}^{z} \frac{dz'}{H(z', \theta)},
\end{equation}
where, $c$ denotes the speed of light and $H \left(z,\theta \right)$ is the model dependent Hubble expansion rate. The dataset provides observed distance moduli, $\mu_{obs}(z_j)$, together with the full covariance matrix $C$, which accounts for both statistical and systematic uncertainties. To compare theoretical predictions with observations, we employ the chi-square estimator as

\begin{equation}
\chi^{2}_{SN} = \Delta \mu^{\top} C^{-1} \Delta \mu,
\end{equation}
where the residuals are defined as
\begin{equation}
\Delta \mu_{j} = \mu_{\text{th}}(z_{j}, \theta) - \mu_{\text{obs}}(z_{j}).
\end{equation}
Equivalently, this can be expressed explicitly as

\begin{equation}
\begin{split}
    \chi^2(\theta) = \sum_{i=1}^{N} \sum_{j=1}^{N} 
    \big[ \mu_{\text{th}}(z_i,\theta) - \mu_{\text{obs}}(z_i) \big] 
    (C^{-1})_{ij} \\
    \times \big[ \mu_{\text{th}}(z_j,\theta) - \mu_{\text{obs}}(z_j) \big],
\end{split}
\end{equation}

with $N=1701$ for Pantheon+SH0ES dataset. Once Hubble, BAO, and Pantheon+SH0ES datasets have been analyzed individually, we may also take the combination of all to determine the best-fit values of our model. To do this, the $\chi^2_{total}$ function can be set up as the sum of $\chi^2_{H(z)}$, $\chi^2_{BAO}$, and $\chi^{2}_{SN}$, as
\begin{equation}
\chi^2_{total}= \chi^2_{H(z)} + \chi^2_{BAO} + \chi^2_{SN}.
\end{equation}
After incorporating the cosmological datasets to parameterize the free parameters of the parametric form, we obtained the best-fit values of the parameters as given in  TABLE--\ref{TABLEI}. We will investigate all the cosmological and dynamical parameters by considering the median values constrained through MCMC. The contour plot generated from $H(z)$, BAO, Pantheon+SH0ES and Combined datasets are respectively presented in FIG.--\ref{MCMC with Hubble dataset}, FIG.--\ref{MCMC with BAO dataset}, FIG.--\ref{MCMC with Pantheon+SH0ES dataset} and FIG.--\ref{MCMC with combined dataset}.

\begin{table*}[htbp] \label{TABLEI}
\centering
\renewcommand{\arraystretch}{1.4}
\setlength{\tabcolsep}{10pt}
\captionsetup{justification=raggedright, singlelinecheck=false}
\caption{Best-fit values and corresponding 1$\sigma$, 2$\sigma$, and 3$\sigma$ uncertainty ranges for the parameters $H_0$, $a$, and $b$ derived from the $H(z)$, BAO, Pantheon+SH0ES, and combined datasets.}
\label{table:confidence_intervals}
\begin{tabular}{llccc}
\toprule
\textbf{Dataset} & & \boldmath{$a$} & \boldmath{$b$} & \boldmath{$H_0$} \\ 
\hline
{$H(z)$} 
 & $1\sigma$ & $0.45^{+0.10}_{-0.09}$ & $1.06^{+0.10}_{-0.09}$ & $72.74^{+1.81}_{-1.79}$ \\ 
 & $2\sigma$ & $0.45^{+0.19}_{-0.19}$ & $1.06^{+0.19}_{-0.18}$ & $72.74^{+3.62}_{-3.50}$ \\ 
 & $3\sigma$ & $0.45^{+0.27}_{-0.28}$ & $1.06^{+0.29}_{-0.28}$ & $72.74^{+5.52}_{-5.27}$ \\ 
\hline
{BAO} 
 & $1\sigma$ & $0.078^{+0.097}_{-0.097}$ & $0.952^{+0.060}_{-0.058}$ & $69.706^{+1.037}_{-1.022}$ \\ 
 & $2\sigma$ & $0.076^{+0.191}_{-0.193}$ & $0.952^{+0.118}_{-0.115}$ & $69.706^{+2.032}_{-2.012}$ \\ 
 & $3\sigma$ & $0.076^{+0.292}_{-0.297}$ & $0.952^{+0.181}_{-0.171}$ & $69.706^{+3.160}_{-3.030}$ \\ 
\hline
{Pantheon+SH0ES} 
 & $1\sigma$ & $0.351538^{+0.018994}_{-0.019226}$ & $1.005345^{+0.049304}_{-0.049537}$ & $74.193858^{+0.135798}_{-0.136570}$ \\ 
 & $2\sigma$ & $0.351538^{+0.038111}_{-0.038376}$ & $1.005345^{+0.097523}_{-0.097784}$ & $74.193858^{+0.272150}_{-0.274277}$ \\ 
 & $3\sigma$ & $0.351538^{+0.056460}_{-0.057528}$ & $1.005345^{+0.147710}_{-0.149641}$ & $74.193858^{+3.041}_{-0.420057}$ \\ 
\hline
{Combined}
 & $1\sigma$ & $0.32933^{+0.03757}_{-0.03935}$ & $1.15460^{+0.05719}_{-0.05681}$ & $74.01162^{+0.20260}_{-0.19831}$ \\ 
 & $2\sigma$ & $0.32933^{+0.07358}_{-0.07939}$ & $1.15460^{+0.11510}_{-0.11254}$ & $74.01162^{+0.40396}_{-0.39481}$ \\ 
 & $3\sigma$ & $0.32933^{+0.10904}_{-0.12189}$ & $1.15460^{+0.16982}_{-0.16791}$ & $74.01162^{+0.59330}_{-0.59186}$ \\ 
\hline
\end{tabular} \label{TABLEI}
\end{table*}

\begin{figure}[htbp] 
   \centering 
   \mbox{\includegraphics[scale=0.5]{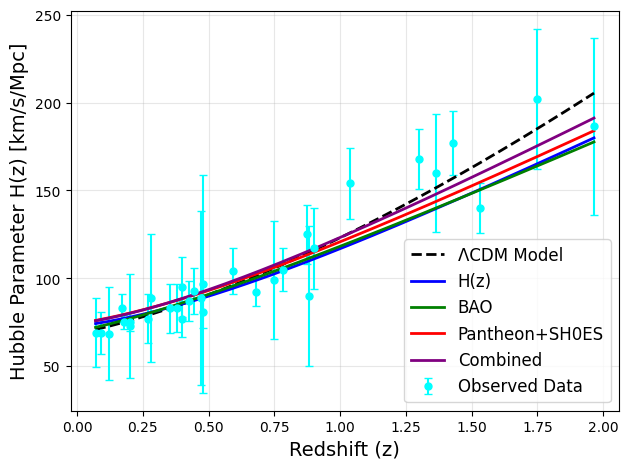}}   
    \hspace{10px}
    \mbox{\includegraphics[scale=0.5]{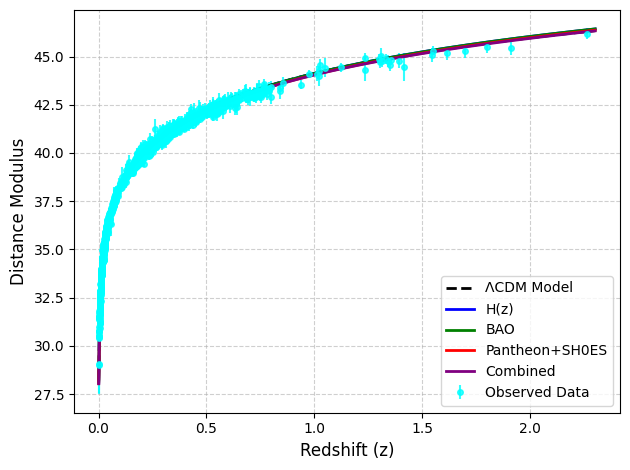}}
    \caption{\raggedright (i) {\bf Upper panel:} Error bar plot of Hubble parameter in redshift;  {\bf Lower panel:} Error bar plot of distance modulus in redshift. The curves are based on the constraints from $H(z)$, BAO, Pantheon+SH0ES and Combined datasets.}
   \label{errorbar}
\end{figure}

\begin{figure}[htbp] 
   \centering 
   \mbox{\includegraphics[scale=0.7]{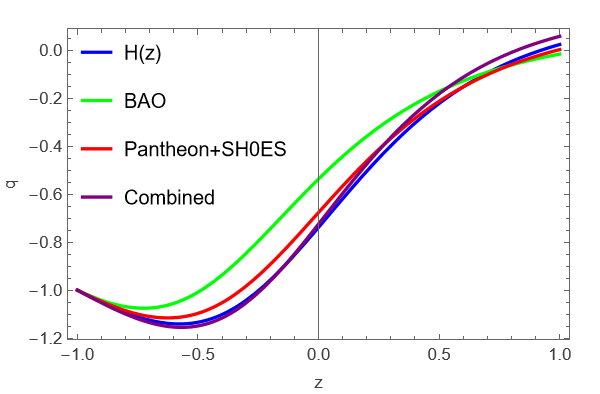}}   
    \caption{ Deceleration parameter in redshift. The curves are based on the constraints from $H(z)$, BAO, Pantheon+SH0ES and Combined datasets.}
   \label{dp}
\end{figure}

The error bar plot of $H(z)$ for all the discussed datasets has been presented presented in FIG.--\ref{errorbar}{\bf(Upper panel)}. It has been observed that the $\Lambda$CDM curve and the model curve are moving well within the error bars. Also, the distance modulus  function $\mu(z)$ of the model for all the dataset is also fitted within the error bar [FIG.--\ref{errorbar}{\bf(Lower panel)}].  

The evolutionary behavior of deceleration parameter for all the discussed datasets is given in FIG.--\ref{dp}. We have observed that the deceleration parameter is decreasing from early time and remains in the negative domain. However, at later times, it increases and settled at $-1$. Though for different datasets its evolution starts from different values however at late times all it merged into a single curve. The present value of the deceleration parameter respectively for  $H(z)$, BAO, Pantheon+SH0ES and combined datasets are $q_0=-0.74$, $q_0=-0.54$, $q_0=-0.68$ and $q_0=-0.73$, which are align with the results obtained in Refs.\cite{NAIK2023138117,Narawade_2024, galaxies11020057,Najafi_2022}.\\

\begin{figure}[htbp] \label{jvss}
   \centering 
   \mbox{\includegraphics[scale=0.7]{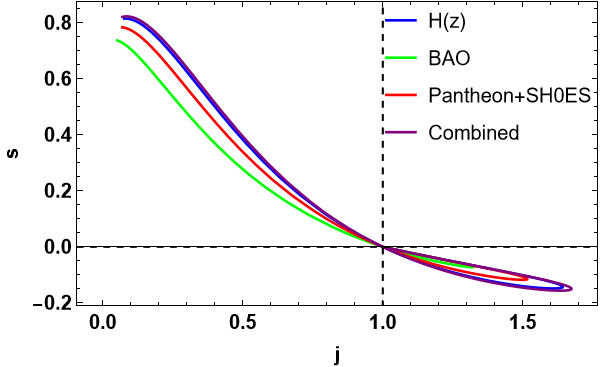}}   
    \caption{Evolutionary behavior of jerk versus snap parameter.  The curves are based on the constraints from $H(z)$, BAO, Pantheon+SH0ES and Combined datasets.} 
   \label{js}
\end{figure}

We shall now present the state finder pair $({j,s})$, a diagnostic tool, which distinguishes various dark energy models. The jerk parameter $j$ and snap parameter $s$ directly derived from the scale factor and are kinematic in nature. Hence, the parameters are dependent only on the characteristics of the metric potentials and are determined by the geometry of spacetime, but not the underlying gravitational theory. These parameters can be obtained as,

\begin{eqnarray}
j(z) &=& q(z)+2 q^2 (z)+(1+z) \frac{d q(z)}{dz}, \nonumber \\
s(z) &=& \frac{j(z)-1}{3 (q(z)-\frac{1}{2})}, \qquad (q \neq \frac{1}{2}). \nonumber
\end{eqnarray}
To note, for $j=1,s=0$, the model appears to be $\Lambda$CDM; for $j<1,s>0$, it leads to quintessence behavior; for $j>1,s<0$ it shows the Chaplygin gas and for  $j=1,s=1$, it indicates $SCDM$. The evolutionary behavior of the $(j,s)$ pair for different datasets is shown in FIG.--\ref{js}. It has been observed that a crossover between the phantom and quintessence phases occur when the trajectories pass through the point $\{1,0\}$ mimicking the $\Lambda$CDM model. 

\section{The cosmological model}\label{model}
To study the dynamical behavior of the cosmological model, we need a functional form of $f(T)$. We consider a well motivated form of $f(T)$ \cite{2011EPJC...71.1797Y}as,
\begin{equation}
f(T) = \frac{\alpha  T_0 \left(\frac{T^2}{T_0^2}\right)^n}{\left(\frac{T^2}{T_0^2}\right)^n+1}+T,
\end{equation}
where $\alpha$, $n$ are model parameters, and $T_0= -6H_0^2$, the present value of torsion scalar. For brevity, we denote the first and second derivatives as  $f_T$ and $f_{TT}$ and can be obtained, respectively as,
\begin{eqnarray}
f_T&=&\frac{2 \alpha  n T_0 \left(\frac{T^2}{T_0^2}\right)^n}{T \left(\left(\frac{T^2}{T_0^2}\right)^n+1\right)^2}+1, \nonumber \\
f_{TT}&=&-\frac{2 \alpha  n T_0 \left(\frac{T^2}{T_0^2}\right)^n \left((2 n+1) \left(\frac{T^2}{T_0^2}\right)^n-2 n+1\right)}{T^2 \left(\left(\frac{T^2}{T_0^2}\right)^n+1\right)^3}. \nonumber
\end{eqnarray}
 Now, Eq. \eqref{rhode}, Eq.\eqref{pde} and Eq.\eqref{omegade} respectively reduce to,

\begin{widetext}
\centering

\begin{equation}
\rho_{de} = \frac{1}{2} \left(-6 \zeta  H_0^2-\alpha  T_0 
\left(1-\frac{36^n (4 n+1) \tau +1}{\left(36^n \tau +1\right)^2}\right)\right),
\end{equation}

\begin{equation}
p_{de} = -\frac{\alpha  H_0^2 \zeta  2^{2 n-1} 3^{2 n+1} T_0 \tau  
\left(8 n^2+1296^n \tau ^2-2^{2 n+1} 9^n (n+1) (4 n-1) \tau -6 n+1\right)}%
{\alpha  36^n n T_0 \tau  \left(-36^n (4 n+1) \tau +4 n-1\right)
-6 H_0^2 \zeta  \left(36^n \tau +1\right)^3},
\end{equation}

\begin{equation}
\omega_{de} = -\frac{\alpha  H_0^2 \zeta  2^{2 n} 3^{2 n+1} T_0 \tau  
\left(8 n^2+1296^n \tau ^2-2^{2 n+1} 9^n (n+1) (4 n-1) \tau -6 n+1\right)}%
{\left(\alpha  36^n n T_0 \tau  \left(-36^n (4 n+1) \tau +4 n-1\right)
-6 H_0^2 \zeta  \left(36^n \tau +1\right)^3\right)
\left(-6 \zeta  H_0^2-\alpha  T_0 
\left(1-\frac{36^n (4 n+1) \tau +1}{\left(36^n \tau +1\right)^2}\right)\right)},
\end{equation}

where $\zeta = (-abz+bz^2+z+1)$ and $\tau = (\frac{H_0^4 \zeta^2}{T_0^2})^{n}$.

\rule{\linewidth}{0.1pt}
\end{widetext}

\begin{figure}[htbp] 
   \centering 
   \mbox{\includegraphics[scale=0.69]{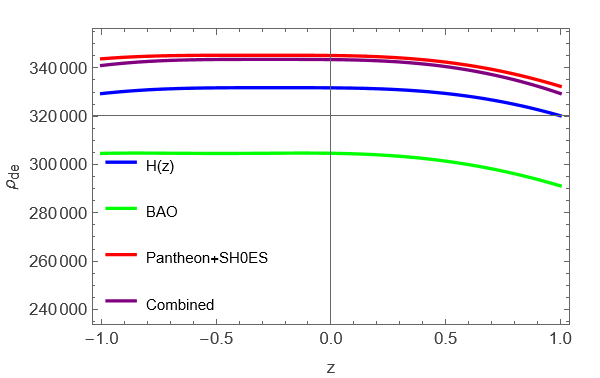}}  
    \hspace{10px}
    \mbox{\includegraphics[scale=0.67]{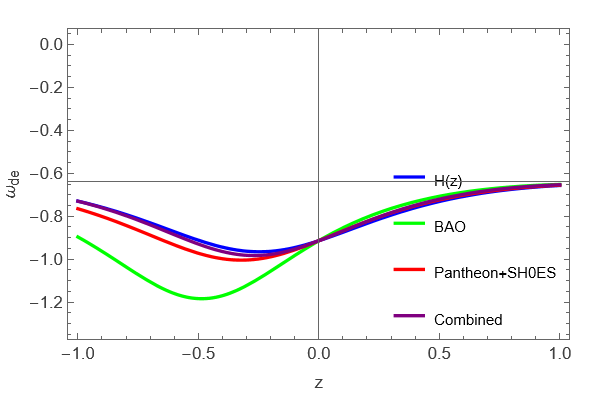}}
    \caption{\raggedright {\bf Upper panel:} The energy density in redshift; {\bf{Lower panel:}} The EoS parameter in redshift. The curves are based on the constraints from $H(z)$, BAO, Pantheon+SH0ES and Combined datasets. The parameter scheme for model parameters are: $\alpha=47.4 $ and $n = 0.038$ .} 
   \label{rhoeos}
\end{figure}
The parameters of the model $a$, $b$, $\alpha$, and $n$ govern the dynamical behavior of the cosmological model. The parameters $a$ and $b$ are constrained with the use of cosmological datasets, whereas the model $\alpha$ and $n$ are fixed to ensure a physically acceptable  energy density. In FIG.--\ref{rhoeos}, the evolutionary behavior of energy density and EoS parameter for the dark sector has been shown. The energy density remains positive throughout the evolution for all the datasets considered. For the datasets $H(z)$, Pantheon+SH0ES and combined case, the EoS parameter remains negative throughout in the range of $[-1,0]$, which represents the quintessence phase. However, in case of BAO dataset, at some part of late times it stays in the phantom region, but ultimately proceed to the quintessence region. At the redshift $z=0$, the value of the EoS parameter are estimated to be $\omega_0=-0.914$, $\omega_0=-0.912$, $\omega_0=-0.915$ and $\omega_0=-0.915$ respectively for $H(z)$, BAO, Pantheon+SH0ES and combined datasets, which are in agreement with some recent Refs.\cite{Duchaniya_2022,doi:10.1142/S0217732322501048}.

Another important point in the modified theory based cosmological model is the evolutionary behavior of energy conditions  \cite{hawking2023large, PhysRev.98.1123,CAPOZZIELLO201899}. These conditions ensure consistency with basic physical principles by placing constraints on the stress-energy tensor. The energy conditions shed light on the nature of matter and energy in the Universe and its possible role in the study of dark energy. They are essentially boundary conditions that influence cosmic evolution \cite{PhysRevD.68.023509}. The properties of gravitational interaction and attraction are also defined by energy conditions, which are crucial because of the basic causal structure of spacetime \cite{doi:10.1142/S0218271819300167}. The energy conditions are  expressed as,
\begin{itemize}
\item{} Null Energy Condition (NEC): $\rho + p \geq 0$;
\item{} Weak Energy Condition (WEC): $\rho \geq 0$, $\rho + p \geq 0$;
\item{} Strong Energy Condition (SEC): $\rho + 3p \geq 0$;
\item{} Dominant Energy Condition (DEC): $\rho - p \geq 0$.
\end{itemize}

\begin{figure*}[htbp]
    \centering
    \setlength{\fboxsep}{6pt} % padding between figure and frame
    \setlength{\fboxrule}{0.8pt} % thickness of the frame border
    \fbox{ % add the box frame
    \begin{minipage}{0.95\textwidth} % slightly less than text width for margin
        \centering
        \begin{minipage}{0.32\textwidth}
            \centering
            \includegraphics[width=\textwidth]{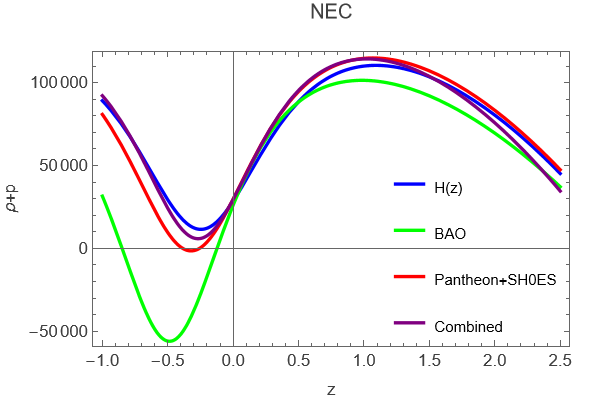}
        \end{minipage}
        \hfill
        \begin{minipage}{0.32\textwidth}
            \centering
            \includegraphics[width=\textwidth]{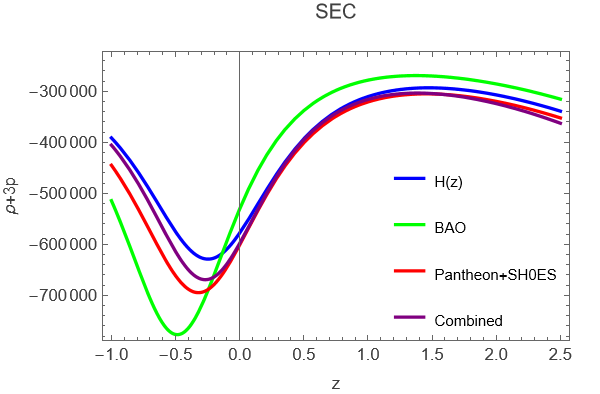}
        \end{minipage}
        \hfill
        \begin{minipage}{0.32\textwidth}
            \centering
            \includegraphics[width=\textwidth]{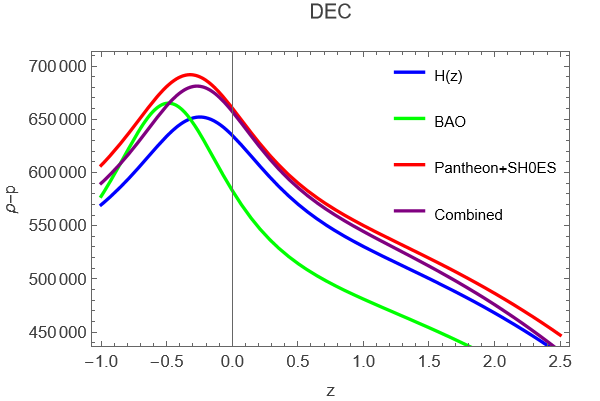}
        \end{minipage}
    \end{minipage}
    }
    \caption{\raggedright Energy conditions in redshift. The curves are based on the constraints from $H(z)$, BAO, Pantheon+SH0ES and Combined datasets. The parameter scheme for model parameters are: $\alpha=47.4 $ and $n = 0.038$.}
    \label{energyconditions}
\end{figure*}

In FIG.--\ref{energyconditions}, the graphical behavior of energy conditions are presented for all the datasets cosnidered. It has been observed that for $H(z)$ and combined datasets, NEC is satisfied everywhere but it is getting violated for BAO and Pantheon+SH0ES in the range around $z \in (-0.85,-0.12)$ and $z \in (-0.38,-0.27)$ respectively. DEC is getting satisfied everywhere for all the datasets. But the SEC is being violated everywhere that supports the accelerating behavior of the Universe.

\section{Conclusion}\label{conclusion}
We have presented an accelerating cosmological model in $f(T)$ gravity with a well motivated form of the functional $f(T)$. The geometrical parameters of the model are analyzed with the constrained values of the Hubble parameterization obtained through the use of cosmological datasets and also with combined datasets. The datasets considered are: $H(z)$, BAO, Pantheon+SH0ES and all combined. The present values of Hubble parameter are in the range $[69.760,74.193858]$. The accelerating behavior has been confirmed from the evolutionary behavior of deceleration parameter and at present, its value remains in the range $[-0.54,-0.74]$. However, at late times, all the curves merged to $-1$. The diagnostic tool, state finder pair has been investigated and the $\Lambda$CDM behaviour has been retrieved.

The dynamical behavior of the model has been assessed through the EoS parameter, which remained negative throughout the evolution. For the $H(z)$, Pantheon+SH0ES and combined datasets, it is lying in the quintessence region, but for BAO, it is lying in the quintessence region at the early time, and just after $z=0$, it is coming to the phantom while moving towards the late time. The estimated current values of the EoS parameter are well within the prescribed range of the cosmological observations. In order to validate the modified gravity theory, the energy conditions are analyzed. It has been seen that SEC violated throughout for all the datasets, whereas DEC is getting satisfied everywhere. The NEC is satisfied everywhere for $H(z)$ and combined datasets but violated for certain redshift values in case of BAO and Pantheon+SH0ES datasets. Nevertheless, this study provides insights in the study of late time cosmic phenomena in the context of modified teleparallel gravity.

\section*{Acknowledgements} BM acknowledges IUCAA, Pune (India) for providing support in the form of an academic visit during which this work is accomplished.

\section{References}
%  \bibliographystyle{utphys}
% \bibliography{references}
% \providecommand{\href}[2]{#2}\begingroup\raggedright

% \endgroup

\end{document}